\documentstyle[preprint,aps,psfig]{revtex}   
\tightenlines
\begin{document}
\draft
\hsize\textwidth\columnwidth\hsize\csname @twocolumnfalse\endcsname
\title{Study of the disordered one-dimensional contact process.}
\author{Raffaele Cafiero $^{1,3}$,  Andrea Gabrielli$^{2,3}$, 
and Miguel A. Mu{\~{n}}oz $^3$}
\address{$^1$  Max-Planck Institut f\"ur Physik komplexer Systeme,
N\"othnitzer strasse 38, D-01187 Dresden, Germany}
\address{$^2$ Dipartimento di Fisica Universit\`a degli Studi ``Tor Vergata'',
v.le della Ricerca Scientifica 1, 00133 Roma Italy}
\address{$^3$
Dipartimento di Fisica e unit\'a INFM,
Universit\'a di Roma ``La Sapienza'',~P.le
A. Moro 2, I-00185 Roma, Italy}
\date{\today }
\maketitle

\begin{abstract}
New theoretical and numerical analysis
of the one-dimensional contact
process with quenched disorder are presented.
 We derive new scaling relations,  different from their 
counterparts in the pure model,  which are
valid not only at the critical point but also away from it
due to the presence of generic scale invariance.
All the proposed scaling laws are verified in numerical simulations.
In addition we map the disordered contact process
into a Non-Markovian contact process
by using the so called
Run Time Statistic, and write down the associated field theory.
This turns out to be in the same universality class
as one derived by Janssen for the quenched system with a Gaussian
distribution of impurities.
 Our findings here
support the lack of universality suggested by the 
field theoretical
analysis:
generic power-law behaviors
are obtained,
evidence is shown of the absence of a characteristic
time away from the critical point, and the absence of universality 
is put forward.
The intermediate
sublinear regime predicted by Bramsom et al. is also found. 

\end{abstract}


\narrowtext

\section{Introduction}
As first conjectured by Janssen and Grassberger \cite{conjecture},
many numerical and analytical studies
have established clearly that all the
systems exhibiting a
continuous  transition
into an {\it unique} absorbing state,  without any other extra
symmetry or conservation law, belong into the same universality class,
namely, that of the contact process (CP) \cite{Kin,Marro}.
That conjecture has been extended to include multicomponent systems
\cite{GG}, and also systems with an infinite number
of absorbing states \cite{inf}.
Among many other models in this broad class are the following:
directed percolation \cite{Kin,Marro}, the contact process \cite{CP},
catalytic reactions on surfaces \cite{catal}, the spreading of epidemics,
and branching annihilating
random walks \cite{bawodd}.
The Reggeon Field Theory (RFT)
is the minimal continuous theory capturing the key features of
this universality class \cite{RFT,conjecture},
(which is often referred to as directed percolation
(DP hereafter) universality class).

  Despite of its theoretical importance, no experiment has 
succeeded so far in 
identifying critical exponents compatible with the predicted DP values.
This could be due to the fact that real systems are never pure, i.e,
they present impurities, dilution or other forms of disorder.
 The question arises 
of how disorder affects the critical behavior of DP-like systems. 
 That problem was first posed by Kinzel
 \cite{Kinzel} and studied numerically
by Noest \cite{Noest1,Noest2}
who showed using
a Harris criterium \cite{Harris} that quenched disorder changes the
critical behavior of DP systems in spatial dimensions below $d=4$.
He also demonstrated that in $d=1$ 
generic power law (generic scale invariance) can be observed, and 
that
 in $d=2$ a Griffiths-like phase \cite{Grif}
 can appear when the impurities take
the form of dilution \cite{Noest1}. This same problem has been
recently tackled by Dickman and Moreira in an interesting
series of papers \cite{ron1,ron2} where they have pointed out the 
presence of logarithmic time-dependences in $d=2$
 and a possible violation of 
scaling.

In any case, the dynamics in impure DP-systems
 is well established to be extremely slow:
due to the presence of impurities, a system that globally is in 
the absorbing phase, can include regions that take locally parameter values 
that correspond to the active regime in the 
analogous pure     
system.
The presence of these zones 
makes it difficult for the system to relax 
to the absorbing state, and consequently 
it decays in a slow fashion: i.e,
 exhibiting
 power laws in $d=1$ \cite{Noest1,Noest2},
 and logarithmicly in $d=2$ \cite{ron1,ron2}, but not
exponentially as generically occurs in pure systems away from the 
critical point.

  The problem of temporally disordered systems with absorbing states
has also been recently investigated, with apparently striking
conclusions \cite{Iwan}.

     At a theoretical level a field-theory analysis for
this class of impure systems has
recently been derived by Janssen \cite{Janssen}. His works corrects
a previous incomplete analysis \cite{Obukhov}, and concludes 
from an epsilon expansion around the upper critical dimension, $d=4$,
 that the
 renormalization group flow equations exhibit only
 runaway
trajectories, and therefore there is no stable (perturbative) 
fixed point (nothing can be concluded about non-perturbative fixed points).
This can be seen as an evidence that no universal critical behavior is 
expected in this class of models.

   In this paper we revisit the impure one-dimensional problem, 
and look at it within a new perspective. In particular, we analyze
the presence or absence of scaling laws in analogy with the two-dimensional 
results recently presented by Dickman and Moreira,
study the universality of critical exponents and the scaling relations
they obey, and verify the presence of a 
sublinear regime predicted by Bramson et al. \cite{Bramson}.
 On the other hand, we present a non-Markovian representation of this
class of systems that shows the same phenomenology, and derive from
it a field theory that turns out to be equivalent to the one derived
by Janssen. From the field theory we obtain new relations among 
exponents.

 \section{The model}

   In the standard contact process \cite{CP,Kin} each site of a d-dimensional 
lattice is either 'occupied' or 'vacant'. In its discrete-time
version, an 
occupied site is extracted randomly at each time step; it generates
 an offspring with probability $p$, or disappears with 
complementary probability 
 $1-p$. The offspring occupies a randomly chosen nearest neighbor:
if it was empty it becomes occupied, while the system remains unchanged
 if the
neighbor was already occupied.
 In the disordered contact process the probability
 $p$ changes from site to site, is fixed in time, 
and is extracted from a distribution $\Pi(p)$. Through this paper 
we consider in particular:
\begin{equation}
\Pi(p, a)= a p^{a -1},
\label{prob1}
\end{equation}
for which
\begin{equation}
<p>= { a \over a +1},
\label{pmed1}
\end{equation}
in this way $a$ acts as a control parameter. For large values of
$a$, the creation probability  is large, 
and the system is in the active phase. Contrarily, for sufficiently
 small values of $p$ the system decays into the absorbing state. 
 We have chosen the previous distribution for technical reasons: it
simplifies the application of the Run Time Statistic \cite{rts}
that we
employ to study the model.

  The central magnitudes, usually considered in this  kind
of systems are of two types: magnitudes measured in 
analysis with homogeneous initial conditions, 
and those measured studying the spreading
of a localized 'seed' into the otherwise empty space \cite{torre}.

  In the first group, we determine the stationary order parameter
(defined as the average density of particles in the stationary state),
$n$, the correlation time, $\tau$, and the correlation length, $\xi$.

  In the second group we study:        
the  total number of occupied sites
in the lattice
(averaged over all the runs including those which have reached the
absorbing state) as a function of time, $N(t)$,
 the overall surviving
probability $P_s(t)$, that is, the probability that the system has not
reached the absorbing state at time $t$, and the mean square distance
of spreading from the origin of the surviving trials as a
function of time, $R^2(t)$ \cite{torre}.

Right at the critical point of pure systems, we have:
\begin{equation}
   N(t)  \propto  t^{\eta},  ~~~~
  P_s(t)  \propto  t^{-\delta}, ~~~~
  R^2(t)  \propto  t^{z},  ~~~ and ~~~ 
n(t)  \propto  t^{-\theta},
\label{expo}
\end{equation}
and at a small distance $\Delta$ from the critical point,
\begin{equation}
 n \propto \Delta^{\beta},  ~~~~
 \tau \propto \Delta^{-\nu_t},  ~~~~
 \xi  \propto \Delta^{-\nu_x}
\label{expo2}
\end{equation}
which define the set of critical exponents we are interested in.
In pure systems the following scaling relations hold:

\begin{equation}
\eta+  \delta + \theta  = d ~ z/2 ,~~~~ \delta = \theta , ~~~~
   z= 2\nu_x/\nu_t, ~~~~ and ~~~~ \theta=\beta/\nu_t 
\label{sca}
\end{equation}
some of 
these expressions have to be modified for the disordered model as we
will show
(see \cite{Marro,conjecture} and \cite{Yuhai} and references therein).

\section{Non-Markovian representation}

 We start our analysis of the model by mapping it into a 
non-Markovian model. The idea of representing a model 
with quenched disorder by means of an effective non-Markovian 
equation, i.e., with memory, including no disorder, is not a new one.
A complete theory for doing so has been developed in \cite{rts}:
it has been named the Run Time Statistic (RTS),
and has proven to be an useful tool in the study of fractals with quenched
disorder
 \cite{qdbm}, and self-organized models with extremal dynamics \cite{rts1}.

The central idea of the  RTS can 
be exemplified by its application to the random random walker (RRW)
 \cite{Matteo}. The RRW is defined in the following way:
 a standard one-dimensional 
random walker is considered, with the only difference that
the probabilities of jumping to the right, $q$, or to the left, $1-q$,
change from site to site, are quenched, and extracted from a certain
probability distribution $P(q)$.
The probability that at a given site, characterized by a given value of 
$q$,
 visited $n$ times by the walker, the walker has jumped $k$
times to the left is given by the binomial distribution:
  \begin{equation}
P(k|q, n) = {\displaystyle n ! \over k !
(n-k)!} q^k (1-q)^{n-k}. 
 \label{binom}
\end{equation}
  Using the Bayes inversion formula for the inversion of
 conditional probabilities one can calculate the
probability that at a given site the probability $q$ takes a particular
  value  between $q$ and $q+dq$ from the knowledge of $k$ after $n$
jumps \cite{Feller,Matteo}
\begin{equation}
P(q+dq |n, k)=  
{\displaystyle (n+1) ! \over k!
(n-k)!} q^k (1-q)^{n-k}  P(q) dq.
\label{pcond}
\end{equation}
 An effective transition probability can be accordingly defined as:
\begin{equation}
q(n,k) = \int dq ~ q  ~  P(q|n,k).
\end{equation}
This equation gives the effective 
probability for the walker to jump
to the right in its $n+1$ visit
to a given
site, conditioned to the fact
that in $n$ previous visits it jumped $k$ times
to the right.

 Observe that the distribution eq.(\ref{pcond}) changes
with time (with $n$); the information about the history of the system
is contained in the effective transition probabilities (that 
change from site to site). This is usually called 
{\it run time statistic} \cite{rts}.

  Let us now apply the previously described method to the disordered
contact process. At each site the value of $p$ (that plays
now a role analogous to $q$ in the RRW), is 
extracted from the  distribution
eq. (\ref{prob1}) \cite{otras}; it is straightforward to verify that 
\begin{equation}
P(p+dp |n, k)= p^{k+a -1} (1-p)^{n-k}
 {\displaystyle (n+1) ! \over (k +a -1) !
(n-k)!} dp
\label{md}
\end{equation}
where now: n is the number of times that a given site has been 
chosen to try an evolution step, and $k$ is the total number 
of times in which an offspring has been generated (obviously
$1-k$ is the number of events in which the site has become empty).
 Therefore the effective parameter $p$ at the site under consideration 
is:
\begin{equation}
<p>= \int_0^1 dp ~ p ~ P(p+dp |n,k) =
{\displaystyle k +a \over n + a +1}.
\label{md2}
\end{equation} 
Note that  the distribution of effective values of the probability
 $<p>$
for any arbitrary $n$ as large as wanted 
does not collapse to a delta function,  
 but converges asymptotically
to the  distribution (\ref{prob1}) \cite{Matteo}.

\section{Field theory}
 
  Using the previously derived non-Markovian approach we can easily
derive an associate field theory.  
 Let us first consider the standard Reggeon field theory describing the
universality class of the pure contact process \cite{conjecture,RFT}.
\begin{equation}
  S[\phi,\psi]=
\int dr^d \int_0^\infty dt \left[ \lambda \psi(x,t)^2 \phi(x,t) -\psi(x,t) 
\left( \partial_t \phi - \mu^2 \phi - \lambda \phi(x,t)^2
-\nabla^2 \phi(x,t) \right) \right].
\label{rft}
\end{equation}
The coefficient of the
  linear term, $\mu^2$ (the {\it  mass}  in a field theoretical language),
depends linearly on the creation probability $p$. A large $p$
renders the contact process supercritical, and so does a value of
$\mu^2$ 
above its critical value.  Observe that at any time the 
renormalized value of $\mu^2$ at a given point $x$
 is given by the expectation value of 
 $ \psi(x,t) \phi(x,t)$.    
 In order to implement the dependence
of p on the history at each point $x$, 
 we can  perform the following substitution:
  \begin{equation}
 \mu^2 \rightarrow \mu_{mod}^2 (x,t) = \mu^2 + \gamma
 \int_0^t d \tau  \psi(x, \tau) \phi(x, \tau) ,
\label{subs}
\end{equation}
that is, at every time step,  the { \it modified} value of the linear
coefficient, $\mu_{mod}^2$, is given by its original  value corrected 
by a time dependent term: the expectation value of 
 $\psi(x,t) \phi(x,t)$ over
the previous history of the system.
 Observe that $\gamma$ acts as a normalization factor.
 In this way the action becomes:
\begin{equation}
   S_{M}[\phi,\psi]= 
\int dr^d \int_0^\infty dt [ \lambda \psi^2 \phi) -\psi
( \partial_t \phi - \mu^2 \phi - \lambda  \phi^2
 -\nabla^2 \phi ) 
+  \gamma  \psi \phi \int_0^t d \tau  \psi(x, \tau) \phi(x, \tau)]
\label{rft2}
\end{equation}
where the dependence of the fields on $x$ and $t$ has been omitted
for the economy of notation.

 On the other hand considering the standard Reggeon Field theory,
eq. (\ref{rft}),
with a site-dependent quenched mass coefficient,  $\mu^2(x)$,
 and a Gaussian
distribution of 'masses' with 
mean $\mu^2$, and variance $<\mu^2(x) \mu^2(x')>= 1/2 f \delta(x-x')$,
one gets after  averaging over the disorder (for what one just has
to perform a Gaussian integral \cite{Janssen}
\begin{equation}
S_d[\phi,\psi]=\int dr^d \left[
 \int_0^\infty dt \left[ \lambda \psi^2 \phi -\psi
( \partial_t \phi - \mu^2 \phi - \lambda \phi^2
 -\nabla^2 \phi ) \right] \right] 
 + f  \left[ \int_0^\infty dt \psi(x,t) \phi(x,t) \right]^2.
\label{J1}
\end{equation}
Observe that eq. (\ref{J1}) coincides with eq. (\ref{rft2}) except
for a time integration limit and the value of the coefficients.
 This difference states that 
the non-Markovian approach reproduces the exact result in the large time
limit. Such a difference between the two field theories
 can be argued to be irrelevant.

 Naive power counting arguments show that all the three non-linearities
in eq. (\ref{rft2})
can be renormalized in $d=4$. 
This result is consistent with the Harris criterium
\cite{Harris}
presented by Kinzel and Noest in \cite{Kinzel,Noest1},
 which states that quenched
spatial disorder
affects the critical behavior of the contact process and 
models in the same universality class below $d=4$.

 The detailed
renormalization procedure of eq. (\ref{J1}) can be found in 
\cite{Janssen}, where it is concluded that no  stable
fixed point exits below $d=4$.  This result is found 
performing an epsilon expansion around the critical dimension,
and is, therefore, valid only in a perturbative sense. 
  The implication of this fact is that either there is a strong 
coupling fixed point, or there is no fixed point at all. This 
latter possibility could reveal the presence of discontinuous transitions
(that have never been observed numerically) or, in any case, 
a lack of universality. More rigorous conclusions are not
available at this point from the field theoretical analysis.

\section{Scaling laws}
 The field theory we have written down can also be used as
a starting point to derive      
scaling relations. 
 From  eq. (\ref{rft}) (or using other standard scaling 
arguments), it is easy to derive that in the active regime:
\begin{equation}
\eta +\delta + \theta = d z/2.
\label{sr1}
\end{equation}
Let us derive it analogous for the impure (non-Markovian) model
 here using simple arguments:
as $N(t)$ is obtained averaging over all the runs,
 it can be written as $N(t) = N_s (t ) P_s(t) + 0* (1-P_s(t))$
where
$N_s(t)$ is the total number of particles calculated 
averaging only over surviving runs. Consequently, one  gets,
$N_s(t) \approx t^{\eta+\delta}$. After creating a perturbation, 
if a growing cluster of occupied sites is generated,
 the radius of such a cluster 
grows as $R \propto t^{z/2}$ and its volume as
$R^d\propto t^{d z/2}$. From the two previous expressions, the density 
of particles inside the cluster goes like $t^{\eta+\delta-dz/2}$. But 
this density inside the cluster scales as $t^{-\theta}$ by definition
of $\theta$, therefore we have obtained eq. (\ref{sr1}).

  The previous expression is valid only at the critical point in the
pure model, where scale invariance is expected. Contrarily in the 
impure model, where generic scale invariance is expected,
 the previous 
argument is valid in all the active phase, in which growing 
clusters are typically generated from localized seeds.  

  On the other hand, in the absorbing phase, 
typically initial seeds are located in locally absorbing regions 
and die out exponentially. However, there is a probability
for the initial seed of 'landing' in a locally active cluster.
When the perturbation gets out of these clusters dies out
exponentially fast. But inside these finite clusters
the local stationary density is reached in a finite time. 
Therefore we can substitute formally $\theta$ by
 zero in eq. (\ref{sr1}) and get:
\begin{equation}
\eta +\delta = d z/2
\label{sr2}
\end{equation}
(note that this does not mean that $\theta$ is zero). 

 On the other hand 
using the symmetry of the Lagrangean under the exchange of the
fields $\phi$ and $\psi$ it is not difficult to get
\begin{equation}
\delta=\theta 
\label{sr3}
\end{equation}
as in the pure model (see \cite{Yuhai}
 for a review of the underlying ideas).
 Observe that the previous symmetry,
present in the Reggeon field theory, is not broken
by  the introduction of the non-Markovian term, i.e. by the quenched
impurities. Therefore:
\begin{equation}
\eta + 2\delta = d z/2 
\label{sr4}
\end{equation}
in the active regime
of our model, as well as in the critical point of the pure model.

  In the active phase, starting from an homogeneous distribution
the system  relaxes to its stationary state also as a power law.
By definition of the active regime the surviving probability 
does not go to zero for large times, and as $P_s(t)$ is a 
monotonously decreasing function of time, we get that 
$\delta=0$ all along the active phase.
Using the scaling relation eq. (\ref{sr3})
we also get $\theta=0$. 
 This simplifies eq. (\ref{sr4}) to 
\begin{equation}
\eta=dz/2. 
\label{sr5}
\end{equation}

   For completeness let us point out that the exponent $\hat{d}$
  calculated  in \cite{Noest1} is easily found to be related
to the exponents we have defined by
\begin{equation}
\hat{d} = 1 + \eta + \delta.
\label{sr6}
\end{equation}
For that it is enough to observe
 that  $\hat{d}$ is the exponent of a time integral of the
total number of particles averaged over the surviving runs.

  Summing up the main conclusions of this section are:
\begin{itemize}
\item In the active phase:
 $\eta=dz/2$ and
 $\delta=\theta =0$.

\item In the absorbing phase:
$\eta+\delta=dz/2$, and $\delta=\theta \neq 0$.
\end{itemize}


   \section{Monte Carlo results}
 We have performed extensive Monte Carlo simulations of the contact
process with quenched impurities distributed according to eq. (\ref{prob1}),
as well as of the associated non-Markovian contact process defined
by eqs. (\ref{md}) and (\ref{md2}).
   Spreading experiments have been performed in lattices large enough
so the occupied region does not reach the system limits. Experiments
started with a random homogeneous initial condition have been 
performed in system sizes up to $L=10^4$ with periodic 
boundary conditions. Most of the results presented
in what follows correspond to $L=10^3$. At every time step a particle
is randomly chosen and the dynamics proceeds in the way explained in the
model definition section;
after each step the time variable $t$ is increased in $1/N(t)$; i.e. when
all the particles are updated once on average, the time increases in one
unit. Simulations are run long enough as to let the system relax to its     
stationary state in the active phase
 ($t \approx 1.6\times10^5$ time steps). 
The different magnitudes are obtained by averaging 
over many independent runs (from $10^2$ for large values of $<p>$, 
where most realizations die at late times and we can easily collect 
a good statistics, to $10^5$ for small values of $<p>$, where 
many realizations die at early times). All the forthcoming discussions are
valid for both the model with quenched disorder and the non-Markovian
equivalent model: the results are identical within the
numerical accuracy.  In fact, as it is shown in Fig. 1, the 
long time distribution of values of $<p>$ in the non-Markovian model
is verified to converge
to the distribution in the quenched model, eq. (\ref{prob1}).
 To avoid repetition we discuss both cases as a whole, and
present figures for both the disordered and the non-Markovian model.

  The main results we have obtained are the following:

 \subsection{Homogeneous initial conditions}
  
   The density of particles $n(t)$ ( $n(t)=N(t)/L$)
 decays in time as shown in Fig.2.
Observe that for large enough values of $<p>$ the curves converge to a
stationary value, that is, their derivative with respect to time  
converges to zero asymptotically. On the other hand for small values 
of $<p>$ the curves decay like power laws with non-universal exponents
that depends on $<p>$; $n(t) \propto t^{-\theta(<p>)}$.
 In Fig.3 we show the asymptotic exponent 
$\theta(<p>)$ as a function of $<p>$, for both the disordered and 
the non-Markovian model. Observe that it decays continuously
from its maximum value in the absorbing state to a very small 
value (compatible with zero), and it is zero in the active regime.
 Observe also the difficulty to locate accurately the critical point. 
Usualy, i.e. in pure systems, $n(t)$ decays exponentially in the
absorbing phase, converges to a constant in the active phase;
and decays as a power law only at the critical point. Consequently
there is a neat criterium to identify the critical point: power 
laws are the hall-mark of criticality.
 In the impure model, instead, the
generic presence of power laws makes the determination of the
critical point a more delicate issue, but at the same time
a more irrelevant one.

  Two possible scenarios are compatible with the data we have obtained: 
in the first one $\theta(<p>)$ is a continuous function
of $<p>$ and the point in which it 'touches' zero for the first 
time corresponds to the critical point. 
  The second possibility is that there is a discontinuous
jump at the critical point; i.e. the curves in Fig. 3
 would not be continuous; that would imply a non-zero value of
$\theta$ at the transition point.
  Even though from our numerics it is not 
possible to resolve the previous
 dilemma, we are tempted to conclude that the first possibility 
is the right one, based on the small values of $\theta(<p>)$ we 
measure in the vicinity of the critical point, and to the fact
that the slopes are always observed to change smoothly with $<p>$,
therefore, no 'jump' is expected to occur. In any case, from the
numerics $\theta$ can be expressed at the critical point 
as $\theta(<p>_c)=0.02 \pm 0.05$, with $<p>_c=0.71 \pm 0.01$ (see below).

   In Fig.4 we plot the asymptotic density $n$ as a function of $<p>$,
 together with a power law fit.
 The best fit is obtained taking
 $<p>_c=0.705$ for the critical effective parameter, and gives 
$\beta=0.29 \pm 0.01$ for both the disordered and the 
non-Markovian model. In Fig. 5 we check the consistence of our assumption 
on $<p>_c$, by representing in a log-log plot 
 $n^{1/\beta}$ as a function of 
$\Delta=<p>-<p>_c$, with $\beta=0.29$. The extrapolation to zero 
of $n^{1/\beta}$ gives $<p>_c=0.71\pm 0.01$, consistent 
with our previous assumption.
 Observe that the value of $\beta$ we find is very different from the one
obtained by Noest for a different distribution of impurities,
$\beta=1.75\pm 0.1$ \cite{Noest1}. We interpret this fact as 
a consequence of the absence of universality predicted by the field
theory analysis.

  From the previous analysis (which agrees perfectly with the theoretical 
predictions) we can extract the following striking conclusion:
as $\beta$ assumes a finite value and $\theta$ is compatible with zero,
using the scaling relation $\theta= \beta / \nu_t$ 
 we get that 
either $\nu_t$ is infinity or takes an extremely large value. Observe
that Noest measured $\nu_t=4.0 \pm 0.5$
 \cite{Noest1} which is an atypically large value. In fact, an 
analogous result has been obtained in the two-dimensional
version of the model \cite{ron2}; Dickman and Moreira showed
that as a matter of fact the exponent $\nu_t$ is not even 
defined. This is a straightforward consequence of the fact that 
the correlation
functions do not decay exponentially in the absorbing phase,
but as power laws, i.e, there is no associated characteristic
time, and therefore $\nu_t$ is undefined (or formally
   $\nu_t = \infty$).

  In order to further explore this issue we have measured 
the two-time correlation functions, $<n(t_0)n(t_0+t)>-<n(t=\infty)>^2$,
  for large times and different values of $<p>$ for the disordered model. 
The results are presented in Fig. 6. First we observe that
in all the cases, i.e. above, below and at the critical point,
  we get power law behaviors, therefore there is no characteristic
time scale. Second, for a fixed value of $<p>$ and varying $t_0$ we observe 
different transient regimes but the asymptotic behavior does not
depend on $t_0$ for large enough times. This is a prove that
the model does not exhibits aging \cite{aging}; therefore even 
though the field theory representing the model is non-Markovian (i.e., 
the 
two-time correlation functions cannot be expressed only as a function of 
the time difference), 
the system relaxes to an aging-free state.

  This analysis can be interpreted as  further supporting 
the guess that $\theta=0$ at the critical point. Otherwise, using
the scaling relations, we
would get a finite $\nu_t$ and consequently an exponential decay 
of the two-time correlation function.

   \subsection{Spreading}

  In Fig. 7,8 and 9 we present the evolution of the magnitudes
defined in  eq. (\ref{expo}) for the spreading experiments, while 
in tables 1 and 2 we give a summary of the values 
of all the scaling exponents for the magnitudes we have studied, including 
the exponent $\theta$, related to homogeneous initial conditions, 
together with a checking  of the scaling relations between the exponents.
 
Observe that the three magnitudes $N(t)$, $P_s(t)$ and $R^2(t)$ present
generic power law decays. In table 1 we show the values of the
associated exponents $\eta$, $\delta$ and $z$ for different values
of $<p>$. Note that all the scaling laws predicted in the previous
section are
 satisfied generically  within
the accuracy limits. 
 In particular, observe that right at the critical point and in the
active phase
we get a value of $\delta$ compatible with $\delta=0$, and therefore 
satisfying the predicted scaling relation: $\theta=\delta=0$.
 As a byproduct we obtain a confirmation of a result obtained
some time ago by Bramson et al. \cite{Bramson}. They demonstrated
that an impure version of the one-dimensional contact process 
exhibits an intermediate phase, i.e., a region in the active phase
in which $R^2$ grows slower that $t^2$. This is 
accordingly called
the  {\it sublinear regime} \cite{note1}.  We observe sublinear growth
in all the active phase: only in the limit $<p>=1$ ($ a \rightarrow
\infty$) we get linear growth: the intermediate 
phase covers the whole active phase.
Therefore the presence of
such a sublinear regime seems to be a generic feature of impure 
one-dimensional systems with absorbing states.
Our results could be compared with those obtained by Noest for
a different impurity distribution. He got, at the critical point,
 $\hat{d}=1.28 \pm 0.03$,
 which
using eq. (\ref{sr6}), implies $\eta+\delta=0.28 \pm 0.03$, to 
be compared with the value $0.25 \pm 0.02$ that we measure 
(Tabb. \ref{tab1}, \ref{tab2}). On the 
other hand for $z$ Noest measured $z=1.44 \pm 0.06$ (using the      
relation $z=2 \nu_x / \nu_t$),
 and we get $ z= 0.58  \pm 0.02 $, indicating
again a high degree of non-universality (Tabb. \ref{tab1}, \ref{tab2}).

    As a last observation we want to point out that the curves 
for $<R^2(t)>$ in the absorbing phase (see Fig. 9) do not seem 
to have reached their stationary value in the time scale under 
consideration. Thus, the values of $z$ we give in  the tables
must be taken carefully, since the error bars on $z$ 
are quite large. In fact, our results seem compatible with $z=0$, 
asymptotically. Observe also, that the combination $\eta+\delta$
in the absorbing phase gives a small exponent that could also
be compatible with zero asymptotically. 

  In any case, all the predicted scaling relations among exponents
are perfectly satisfied at, above and below the critical point.

\section{Conclusions and summary}
 
  We have studied under different perspectives the disordered
contact process. First we have mapped it into a pure model
with memory, that reproduces all the phenomenology of the original 
 model.     From this new non-Markovian model we write down a simple field 
theory, that is in the same universality class as one presented 
previously for the model with Gaussian-distributed quenched disorder.
Using the field theory we
 have derived a set of scaling relations not only at
the critical point but also in the active and  absorbing phases 
where scale invariance is also observed.
Our theoretical predictions are
 confirmed in extensive Monte carlo simulations:
in particular we have shown the equivalence of the disordered and
the non-Markovian model, the generic presence of scale invariance,
the existence of a sublinear-growth regime, verified the absence
of a characteristic time scale,  and verified all the predicted
scaling laws.  

In a future work we plan to further exploit the mapping into the
non-Markovian model to obtain some other analytical results.
In particular, we pretend to apply real-space renormalization methods
to the one-dimensional model, and try to understand 
the generic scale invariance from a 
renormalization  perspective.

\section{Acknowledgements}
It is a pleasure to
 acknowledge useful discussions with Ron Dickman.
This work has been partially
supported by the European Union through a grant to M.A.M.
(contract ERBFMBICT960925).

\begin{table}
\begin{centering}
\begin{tabular}{|c|c|c|c|c|c|c|c|} \hline
$p$ & $\eta$ & $\delta$ & $z$ & $\theta$ & $\eta+\delta-dz/2$  
& $\eta-dz/2$ & $\delta-\theta$ \\ \hline

$0.5$ & $-0.52\pm 0.02$ & $0.61\pm 0.02$ & $0.12 \pm 0.07$ & 
$0.57\pm0.02$ & $0.03 \pm 0.07$  
& $-$ & $0.04 \pm 0.04$ \\ \hline

$0.55$ & $-0.32\pm 0.02$ & $0.48 \pm 0.02$ & $0.14 \pm 0.07$ & 
$0.44 \pm 0.02$ & $0.09 \pm 0.07$  
& $-$ & $0.04 \pm 0.04$ \\ \hline

$0.66$ & $-0.03\pm 0.01$ & $0.10\pm 0.01$ & $0.16 \pm 0.06$ & 
$0.13\pm0.02$ & $-0.01 \pm 0.04$  
& $-$ & $-0.03 \pm 0.03$ \\ \hline

$0.70$ & $0.19\pm 0.02$ & $0.05\pm 0.01$ & $0.57 \pm 0.02$ & 
$0.02 \pm 0.05$ & $-0.04 \pm 0.04$  
& $-$ & $0.03 \pm 0.06$ \\ \hline

$0.71$ & $0.25\pm 0.02$ & $0.0 \pm 0.01$ & $0.58\pm0.02$ & 
$0.02 \pm 0.05$ & $-$  
& $-0.04 \pm 0.03$ & $-0.02 \pm 0.06 $ \\ \hline

$0.725$ & $0.35\pm 0.02$ & $0$ & $0.72\pm0.02$ & $0$ & 
$-$  
& $-0.01 \pm 0.03$ & $0$ \\ \hline

$0.75$ & $0.53\pm 0.02$ & $0$ & $1.10\pm0.01$ & $0$ & 
$-$  
& $-0.02 \pm 0.03$ & $0$ \\ \hline

$0.8$ & $0.92\pm 0.02$ & $0$ & $1.79\pm0.01$ & $0$ & 
$-$  
& $0.02 \pm 0.03$ & $0$ \\ \hline

$0.85$ & $0.99\pm 0.02$ & $0$ & $1.99\pm0.01$ & $0$ & 
$-$  
& $-0.01 \pm 0.03$ & $0$ \\ \hline

$0.95$ & $1.00\pm 0.02$ & $0$ & $2.00\pm0.01$ & $0$ & 
$-$  
& $0.0 \pm 0.03$ & $0$ \\ \hline
\end{tabular}
\caption{Values of the scaling
 exponents for different values of $p$ (disordered model).}
\label{tab1}
\end{centering}
\end{table}

\begin{table}
\begin{centering}
\begin{tabular}{|c|c|c|c|c|c|c|c|} \hline
$p$ & $\eta$ & $\delta$ & $z$ & $\theta$ & $\eta+\delta-dz/2$  
& $\eta-dz/2$ & $\delta-\theta$ \\ \hline

$0.5$ & $-0.50\pm 0.02$ & $0.54\pm 0.02$ & $0.10 \pm 0.08$ & 
$0.57\pm0.02$ & $-0.01 \pm 0.08$  
& $-$ & $-0.03 \pm 0.04$ \\ \hline

$0.55$ & $-0.37\pm 0.02$ & $0.43 \pm 0.02$ & $0.12 \pm 0.08$ & 
$0.43 \pm 0.02$ & $0.0 \pm 0.08$  
& $-$ & $0.0 \pm 0.04$ \\ \hline

$0.66$ & $-0.08\pm 0.02$ & $0.10\pm 0.01$ & $0.14 \pm 0.07$ & 
$0.11\pm0.02$ & $-0.05 \pm 0.06$  
& $-$ & $-0.01 \pm 0.03$ \\ \hline

$0.70$ & $0.19\pm 0.02$ & $0.03\pm 0.01$ & $0.57 \pm 0.02$ & 
$0.02 \pm 0.04$ & $-0.06 \pm 0.04$  
& $-$ & $0.01 \pm 0.05$ \\ \hline

$0.71$ & $0.24\pm 0.02$ & $0.0\pm 0.01$ & 
$0.58\pm0.02$ & $0.01 \pm 0.04$ & $-$  
& $-0.05 \pm 0.03$ & $-0.01 \pm 0.05$ \\ \hline

$0.725$ & $0.34\pm 0.02$ & $0$ & $0.81\pm0.04$ & $0$ & 
$-$  
& $-0.06 \pm 0.04$ & $0$ \\ \hline

$0.75$ & $0.47\pm 0.02$ & $0$ & $1.03\pm0.02$ & $0$ & 
$-$  
& $-0.04 \pm 0.03$ & $0$ \\ \hline

$0.8$ & $0.93\pm 0.02$ & $0$ & $1.79\pm0.01$ & $0$ & 
$-$  
& $0.04 \pm 0.02$ & $0$ \\ \hline

$0.85$ & $0.99\pm 0.02$ & $0$ & $1.96\pm0.01$ & $0$ & 
$-$  
& $0.01 \pm 0.02$ & $0$ \\ \hline

$0.95$ & $1.00\pm 0.02$ & $0$ & $2.00\pm0.01$ & $0$ & 
$-$  
& $0.0 \pm 0.02$ & $0$ \\ \hline
\end{tabular}
\caption{Values of the scaling exponents for different values of $p$ 
(non-Markovian model).}
\label{tab2}
\end{centering}
\end{table}

\begin{figure}
\caption{ Distribution of $<p>$ for large times
 in the non-Markovian
model $\Pi_M(p)$ compared with the fixed distribution of the
disordered model $\Pi_Q(p)$  for $a=2$.
}
\end{figure}

\begin{figure}
\centerline{\psfig{figure=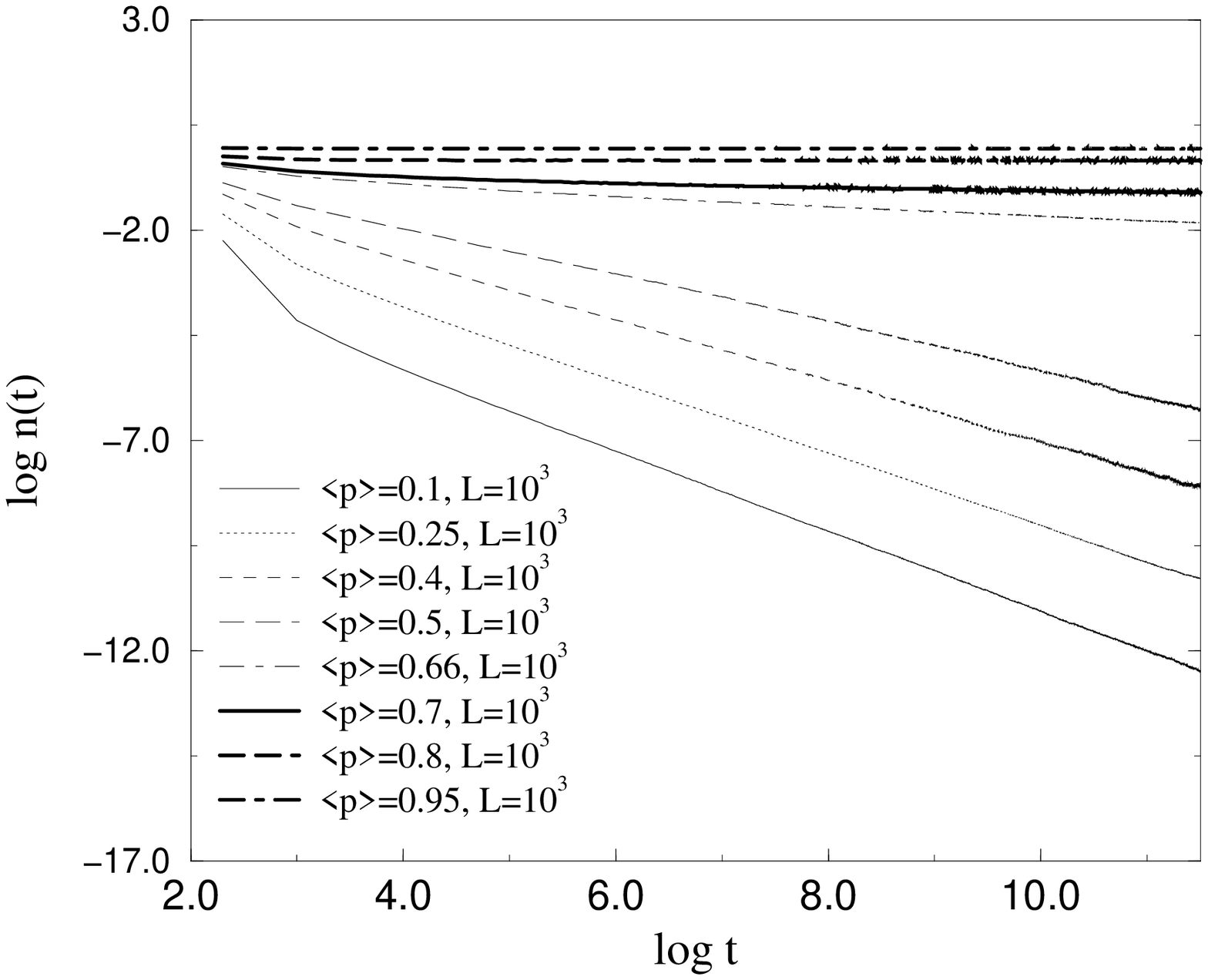,width=6cm}}
\centerline{\psfig{figure=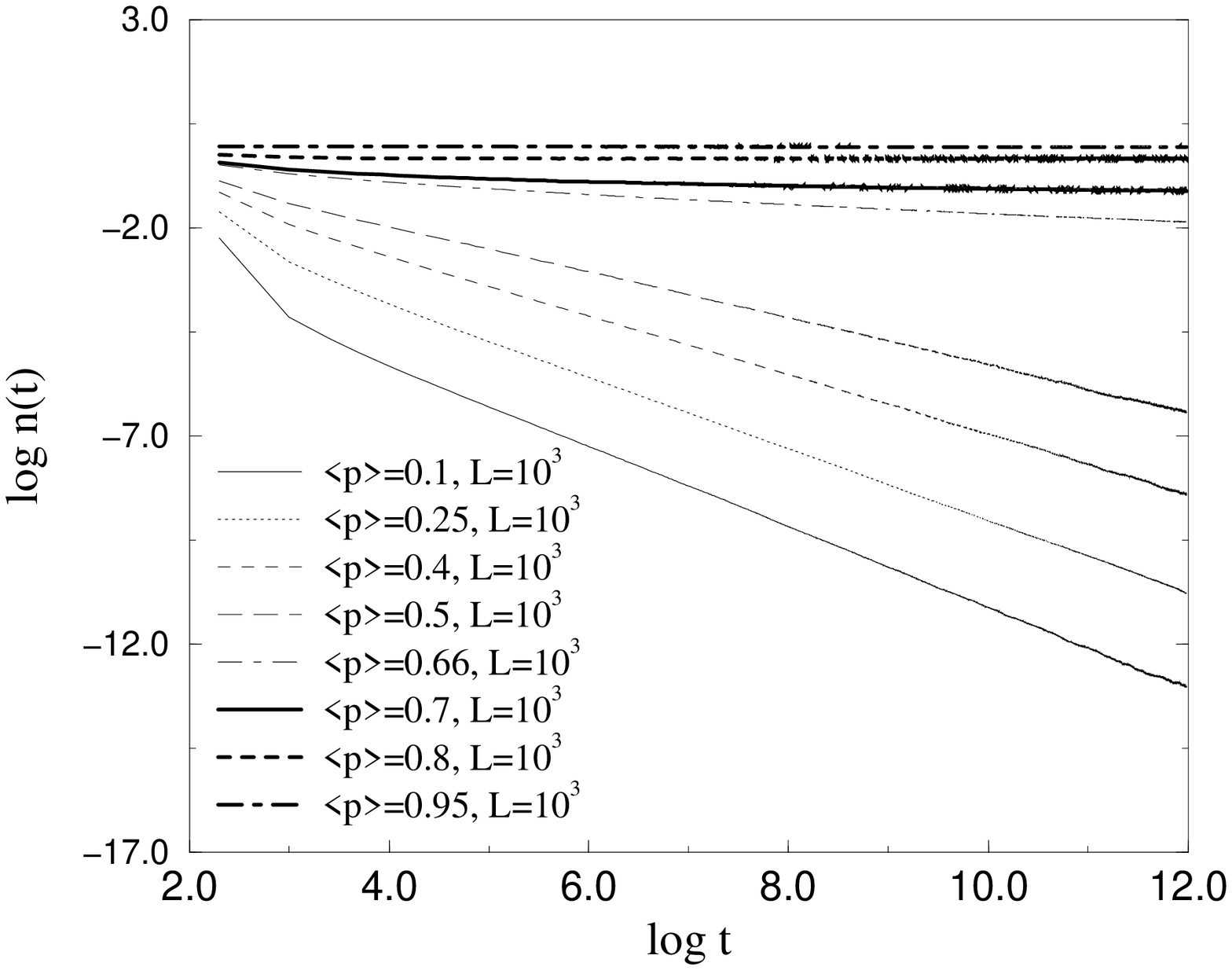,width=6cm}}
\caption{ Decay  of the density of occupied sites 
for different values of $<p>$
 as a function of
time for the disordered model (upper plot), and for the 
non-Markovian model (lower plot).
}
\end{figure}

\begin{figure}
\centerline{\psfig{figure=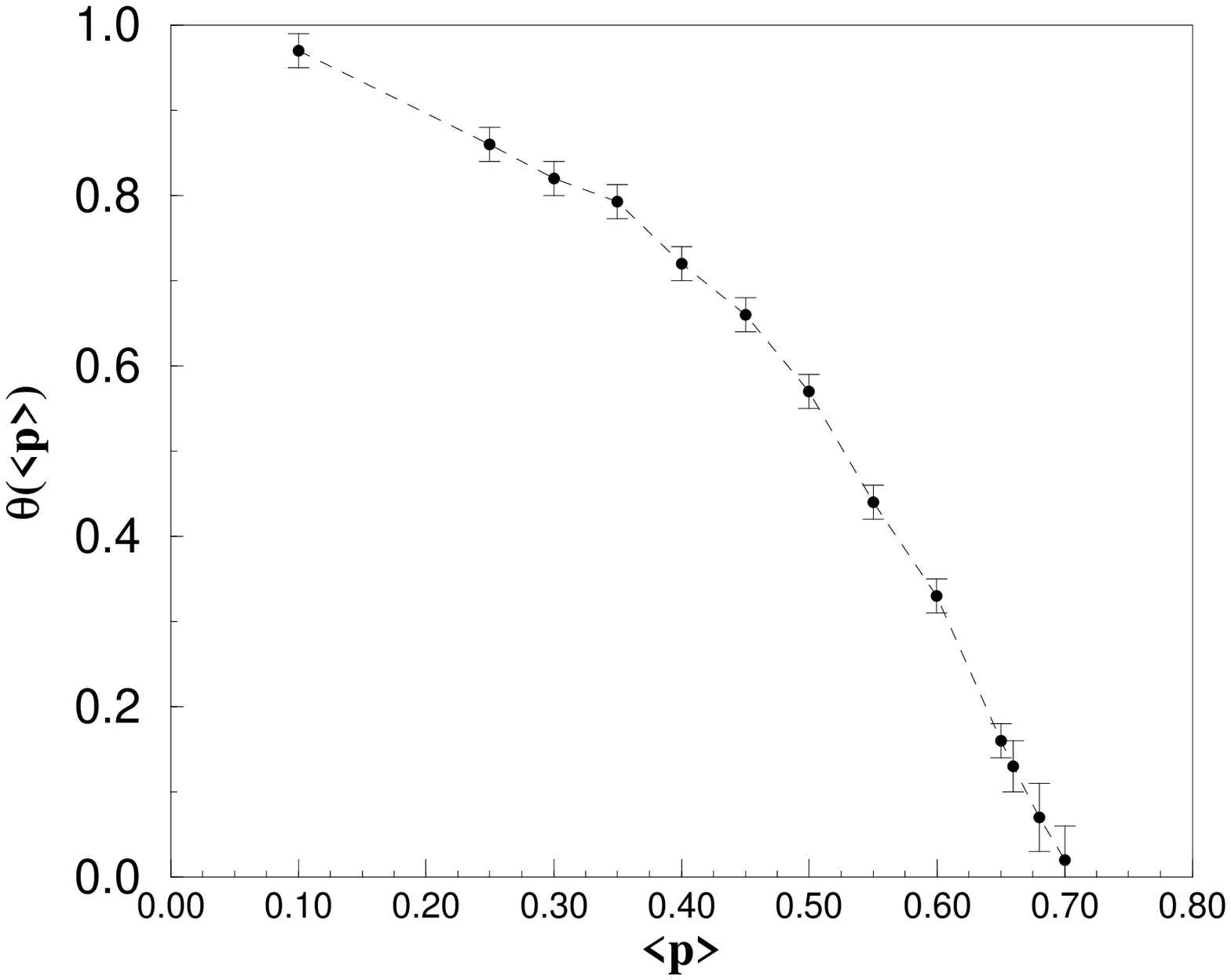,width=6cm}}
\centerline{\psfig{figure=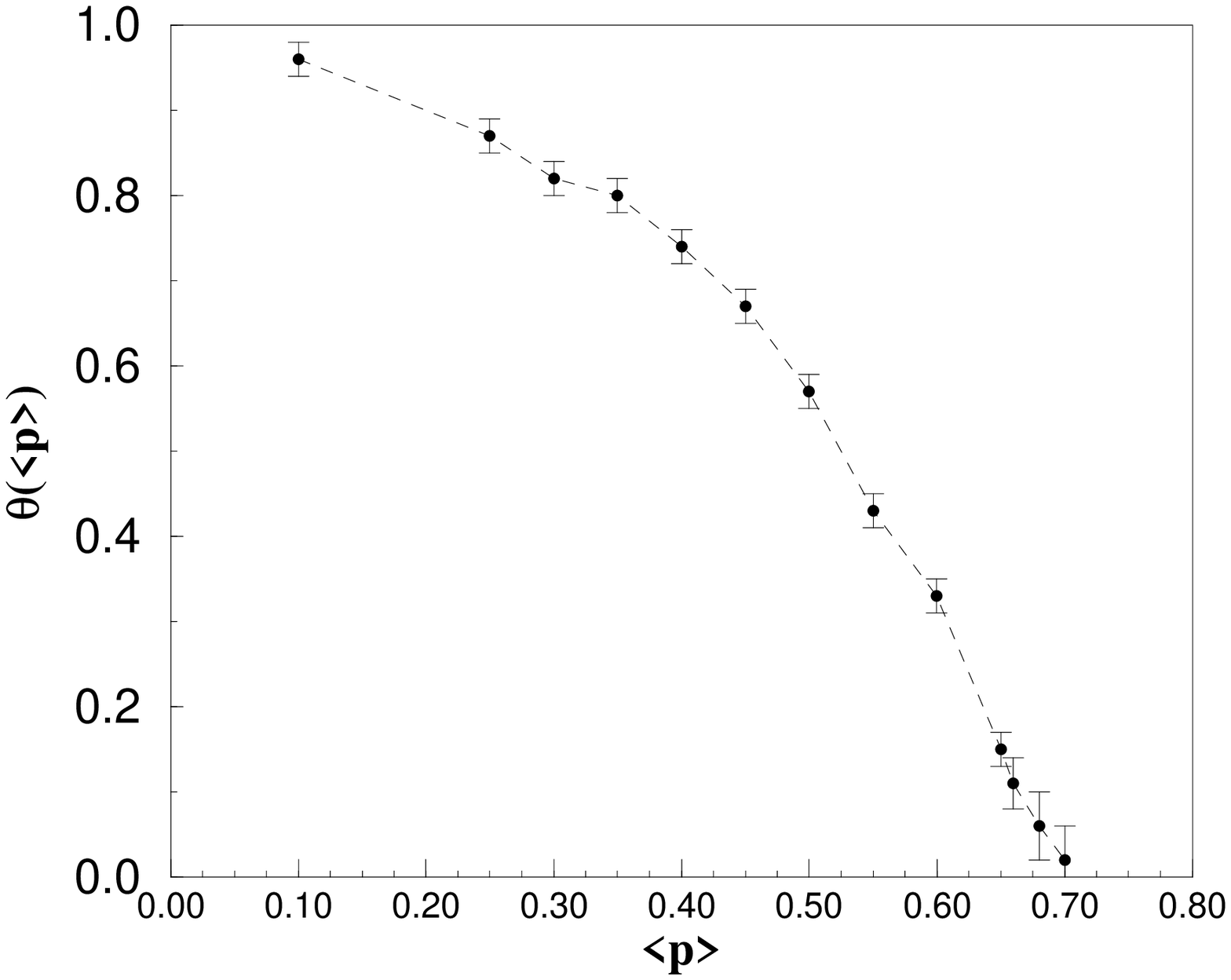,width=6cm}}
\caption{ Value of the exponent $\theta (<p>) $ as a function of $<p>$ for
the disordered model (upper plot), and for the non-Markovian model 
(lower plot).}
\end{figure}

\begin{figure}
\centerline{\psfig{figure=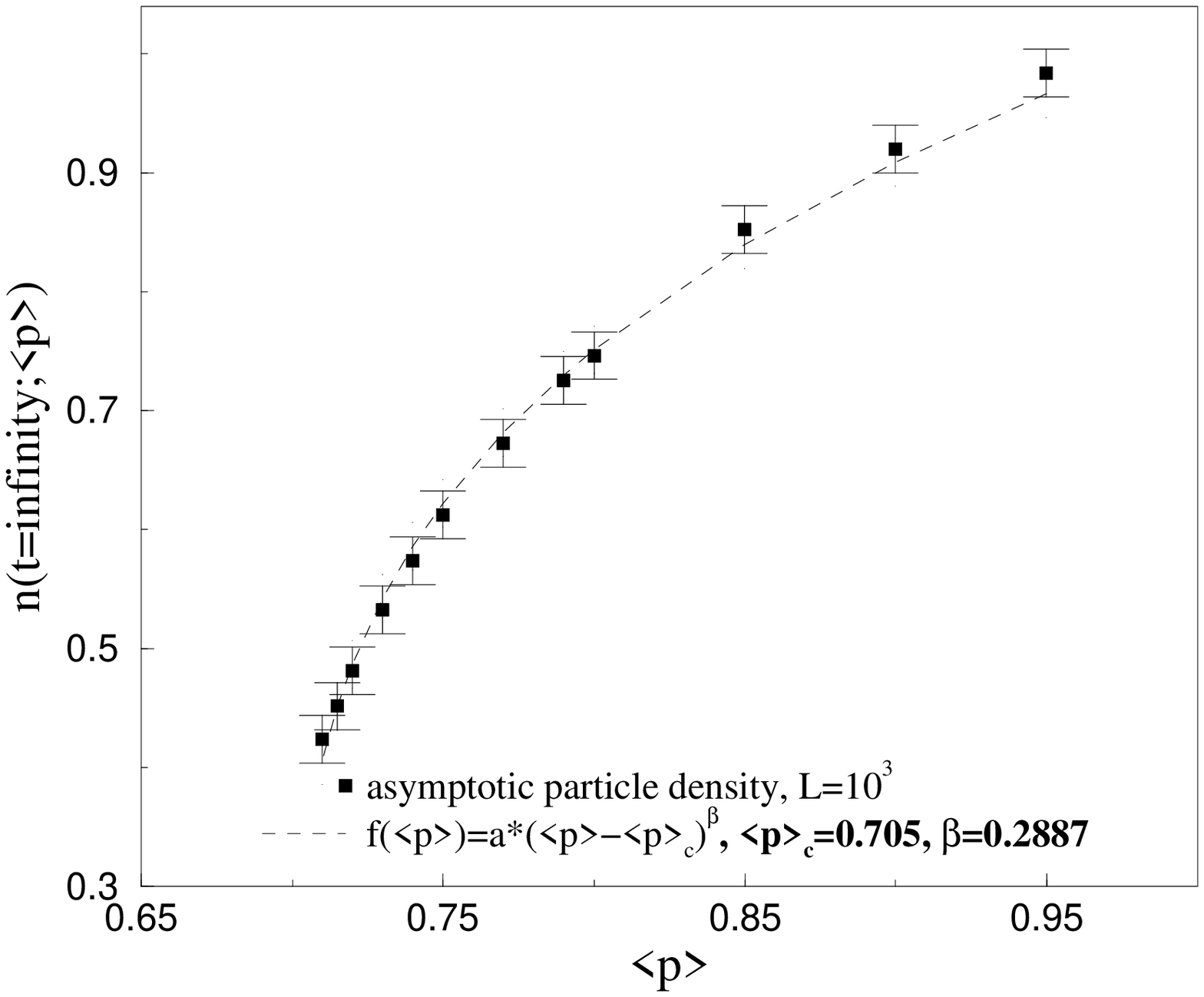,width=6cm}}
\centerline{\psfig{figure=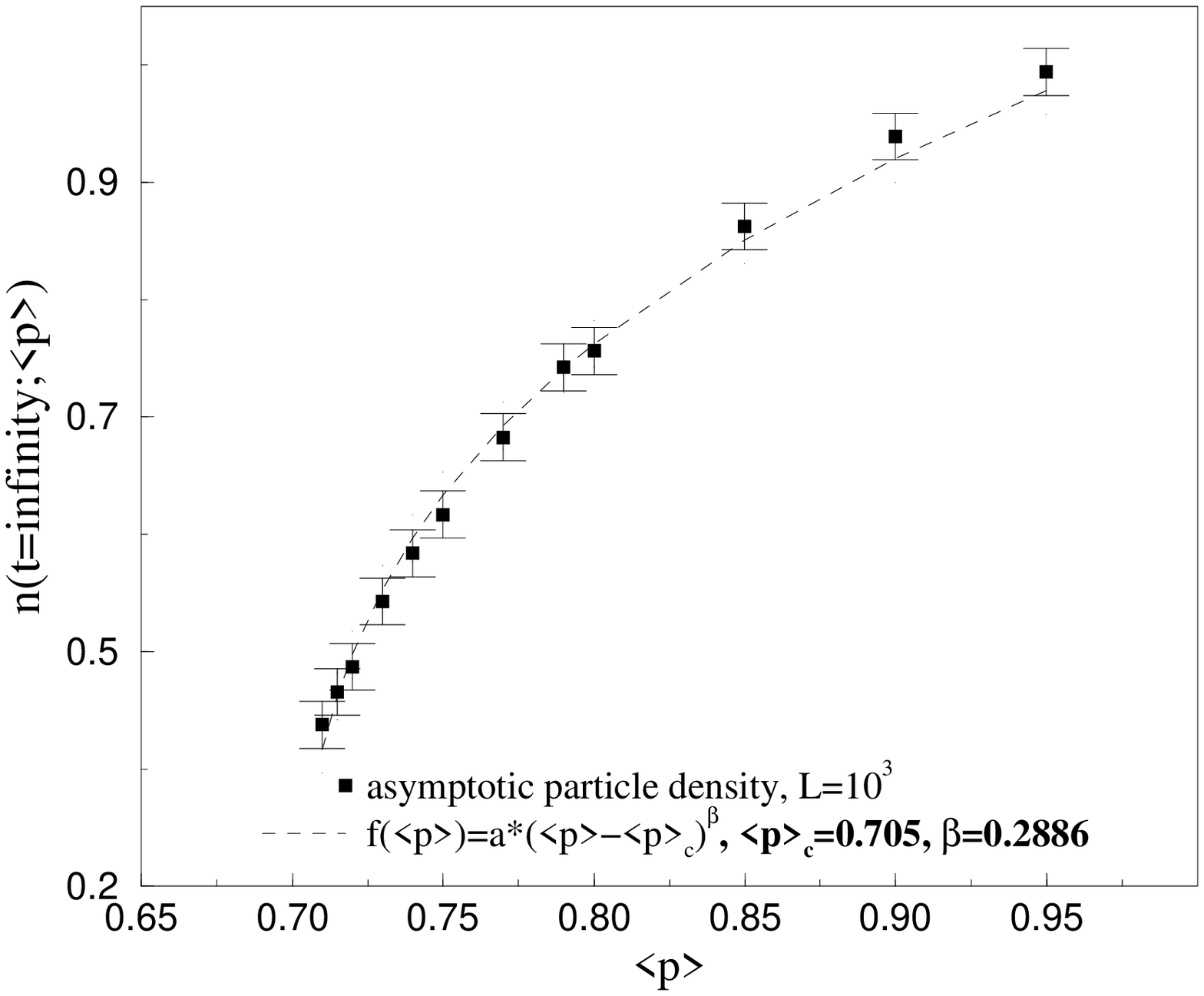,width=6cm}}
\caption{ Stationary value of the density as a function of 
 $<p>$ for the disordered model (upper plot) and for the
non-Markovian model (lower plot). A power law fit for the 
scaling of $n(t=\infty;<p>)$ is shown in the figure.
}
\end{figure}

\begin{figure}
\centerline{\psfig{figure=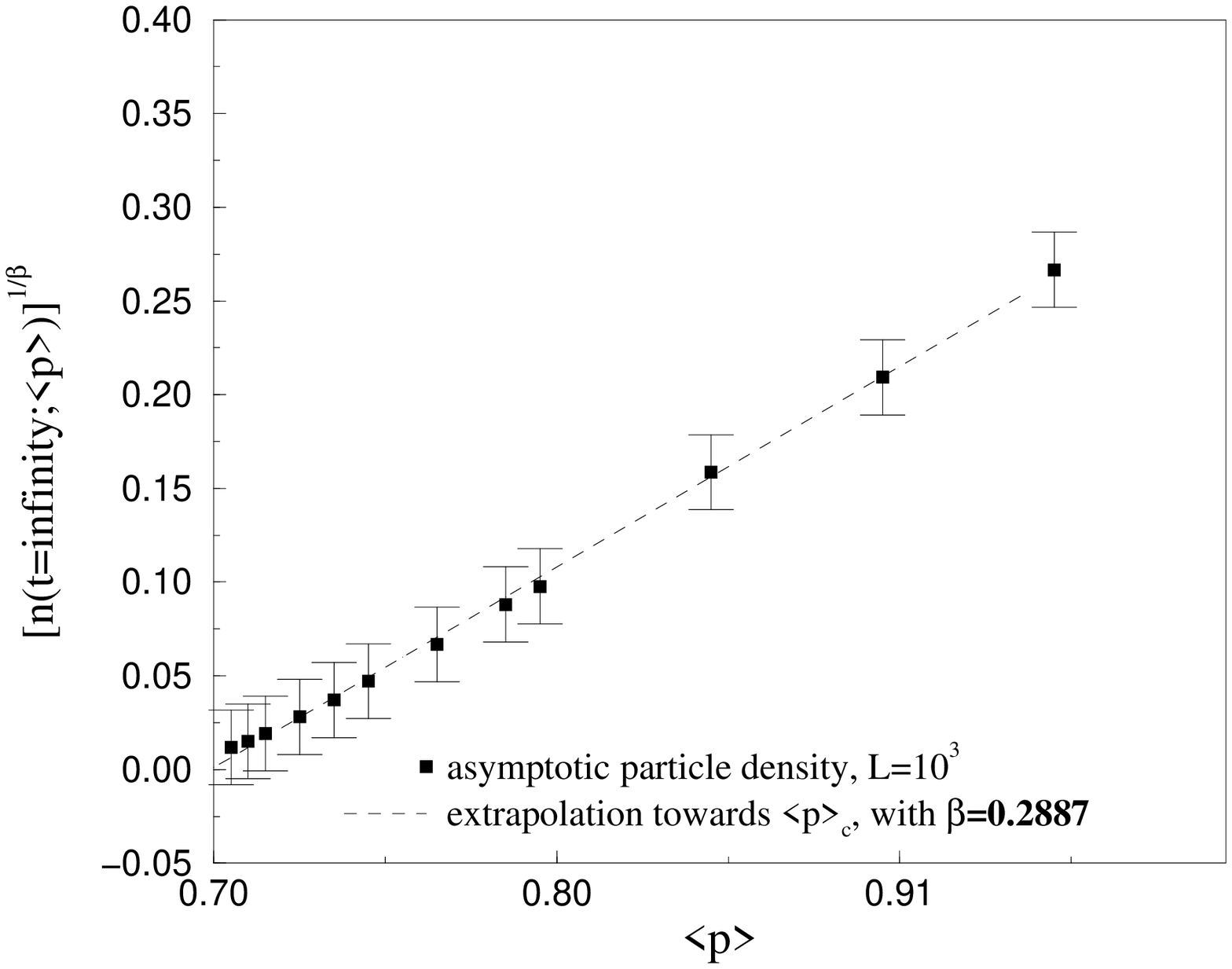,width=6cm}}
\centerline{\psfig{figure=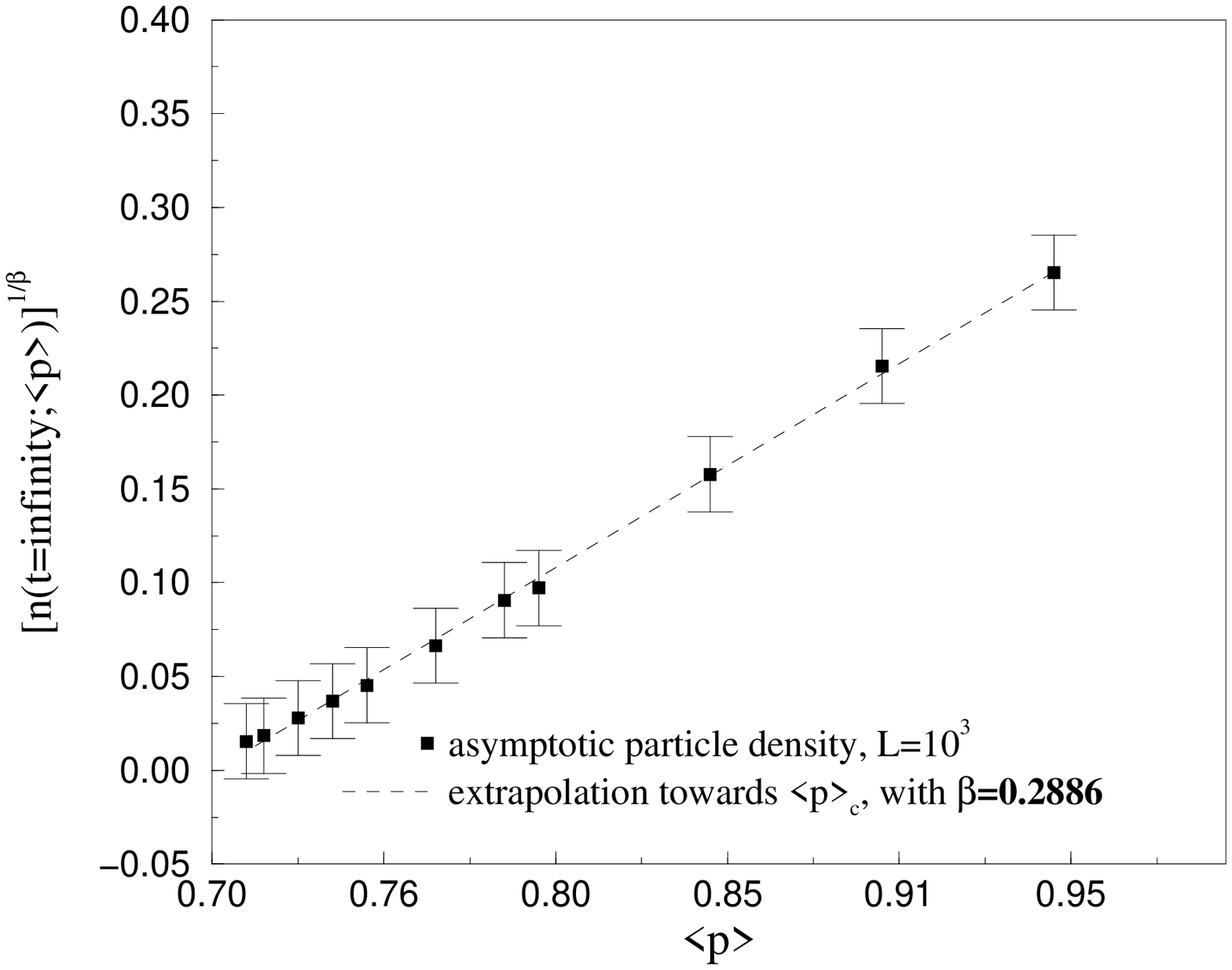,width=6cm}}
\caption{ Log-log plot of
 $n(t=\infty;<p>)^{1/\beta}$ as a function of
 $<p>$ for the disordered model 
(upper plot) and for the
non-Markovian model (lower plot).
 The value of $\beta$ is that given by 
the fit in Fig. 4. The extrapolation to
 $n(t=\infty)=0$ gives $p_c=0.71\pm0.01$.
}
\end{figure}

\begin{figure}
\centerline{\psfig{figure=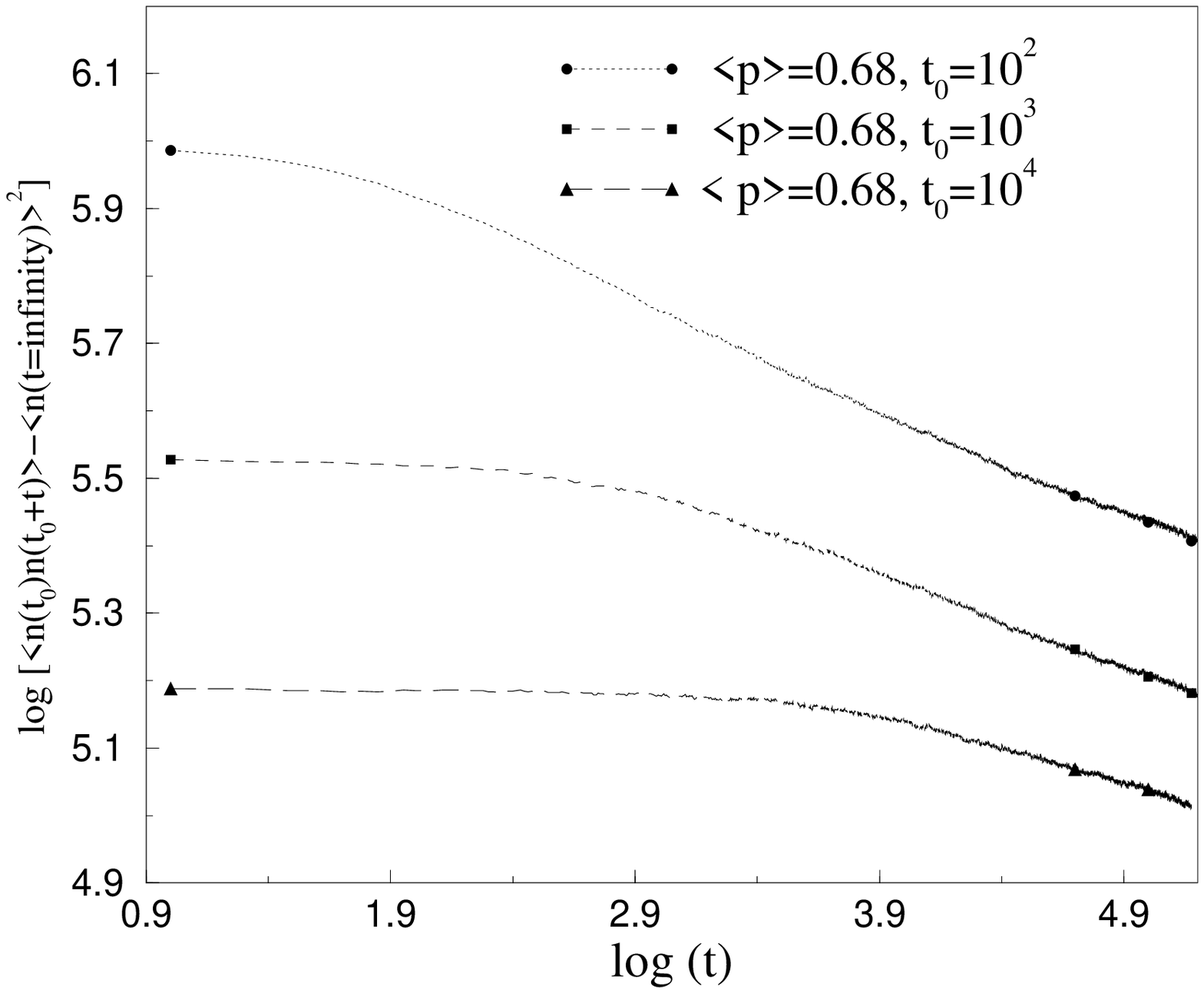,width=6cm}}
\centerline{\psfig{figure=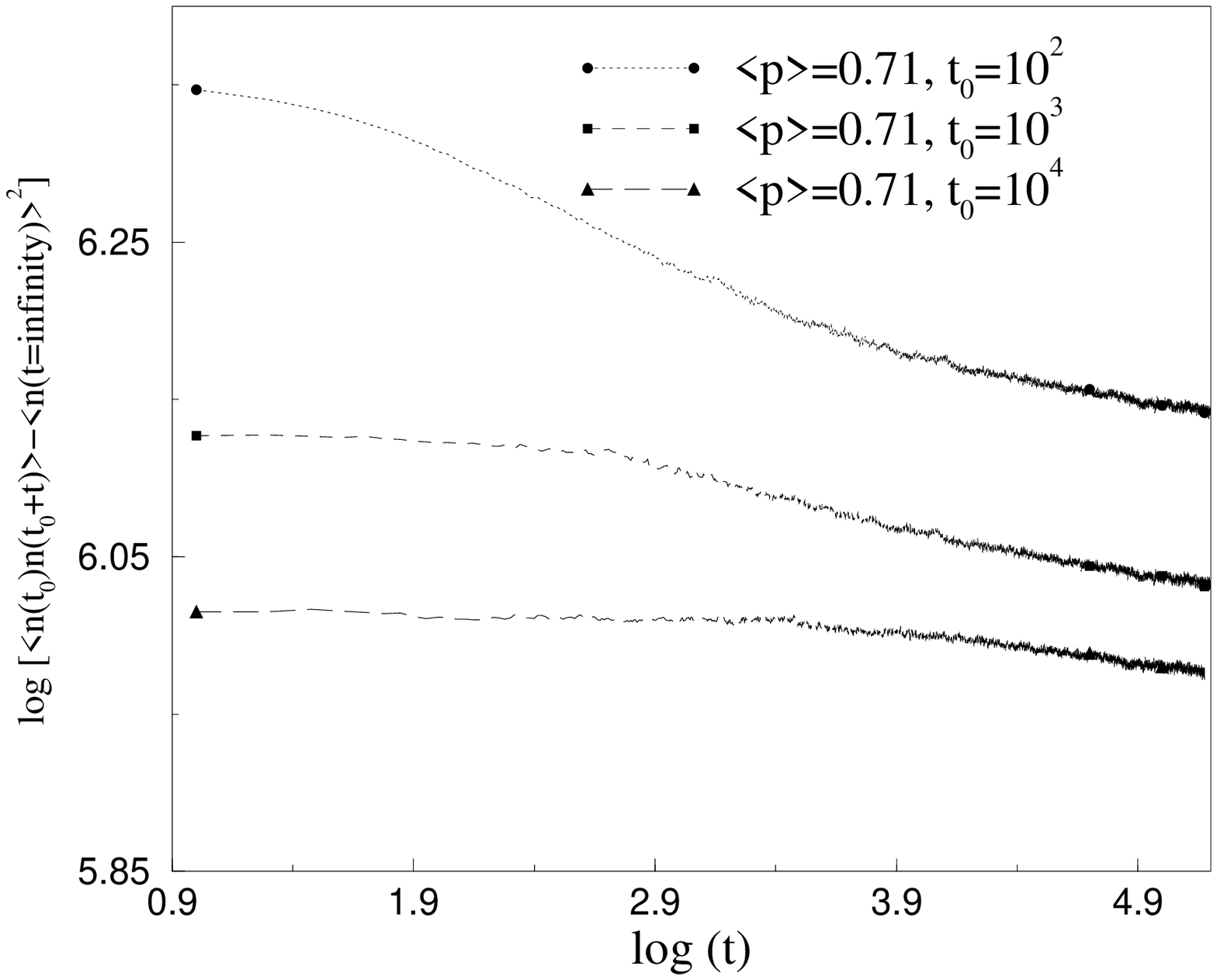,width=6cm}}
\centerline{\psfig{figure=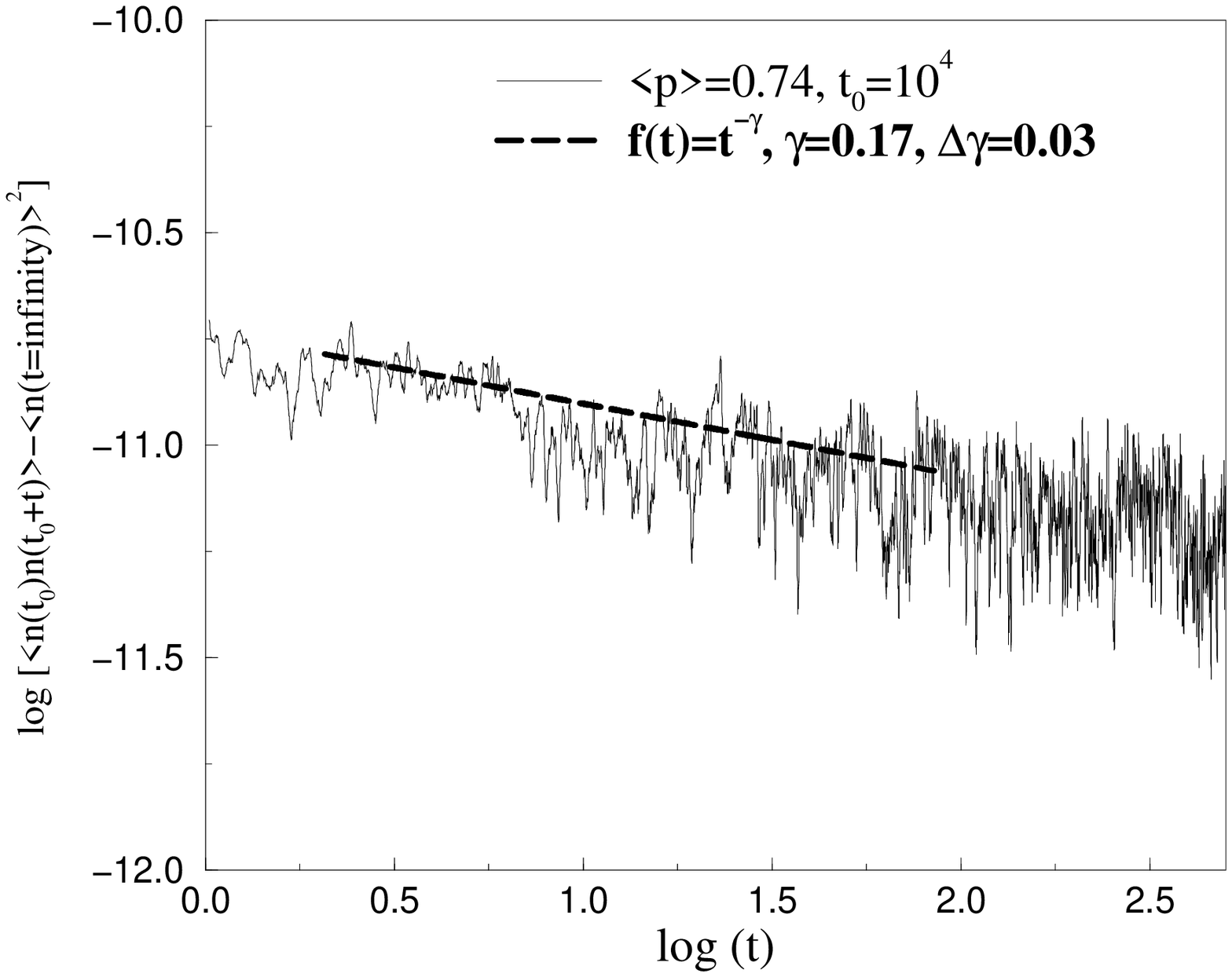,width=6cm}}
\caption{Decay of the two-time density correlation function  for
different initial times (disordered model)
 as a function of the time difference $t$, 
and different values of $<p>$:
 $<p>=0.68$ in the absorbing phase (upper figure),
$<p>=0.71$ at critical point (central figure),
 and $<p>=0.74$ in the active
phase(lower figure).
}
\end{figure}


\begin{figure}
\centerline{\psfig{figure=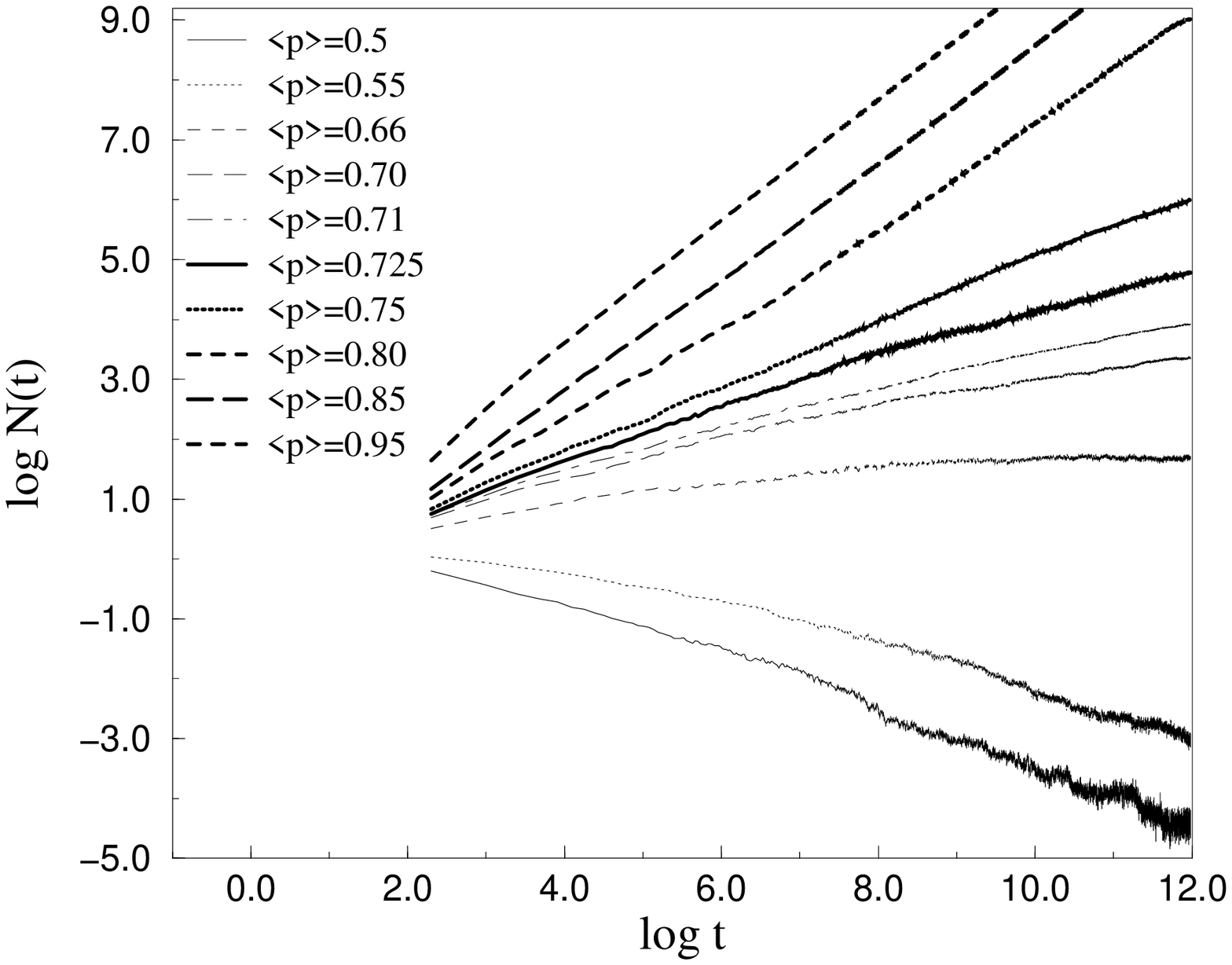,width=6cm}}
\centerline{\psfig{figure=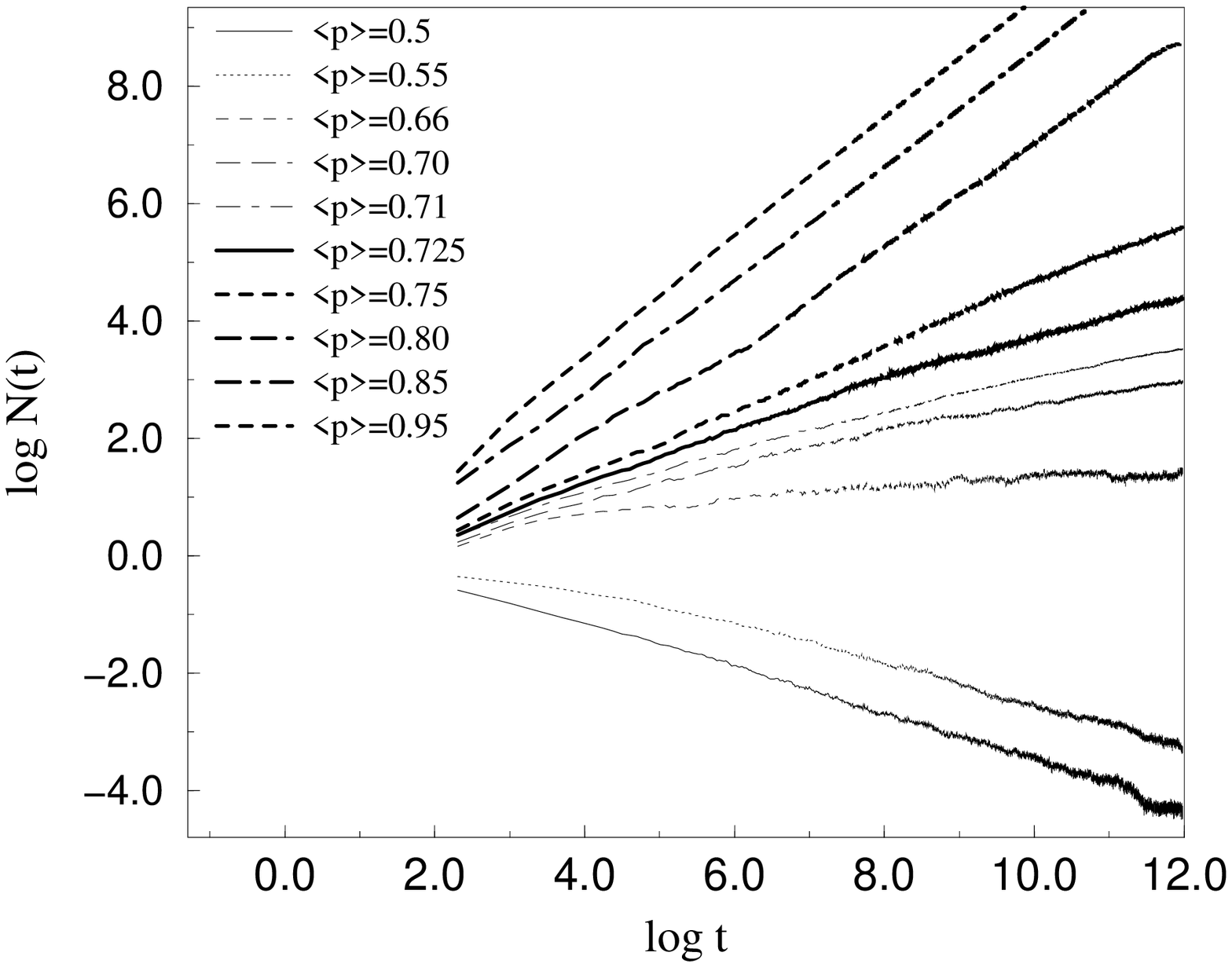,width=6cm}}
\caption{ Averaged total number of particles
 for spreading experiments, with
different values of $<p>$, 
as a function of time for the disordered model (upper plot), and for the
non-Markovian model (lower plot). 
}
\end{figure}
\newpage

\begin{figure}
\centerline{\psfig{figure=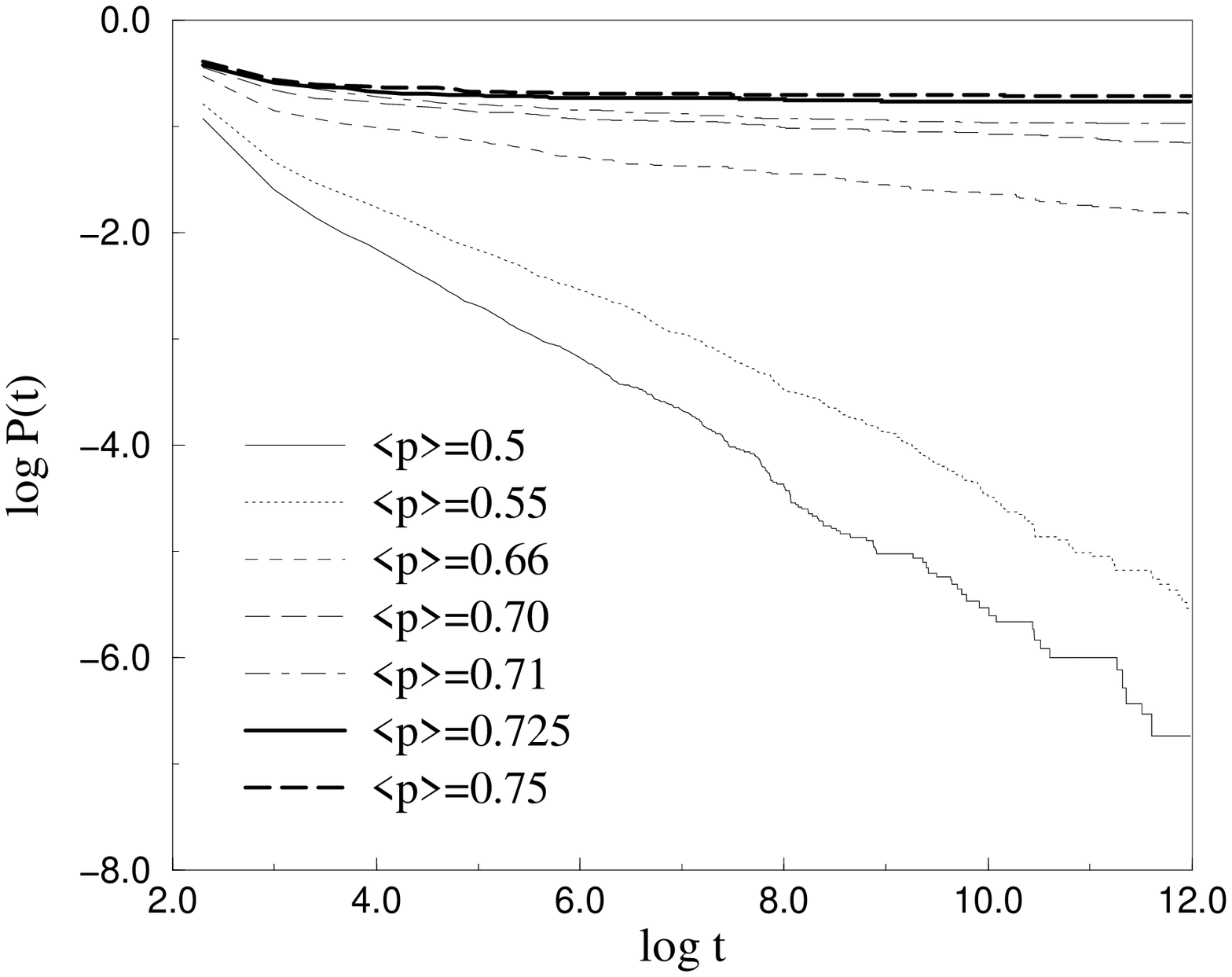,width=6cm}}
\centerline{\psfig{figure=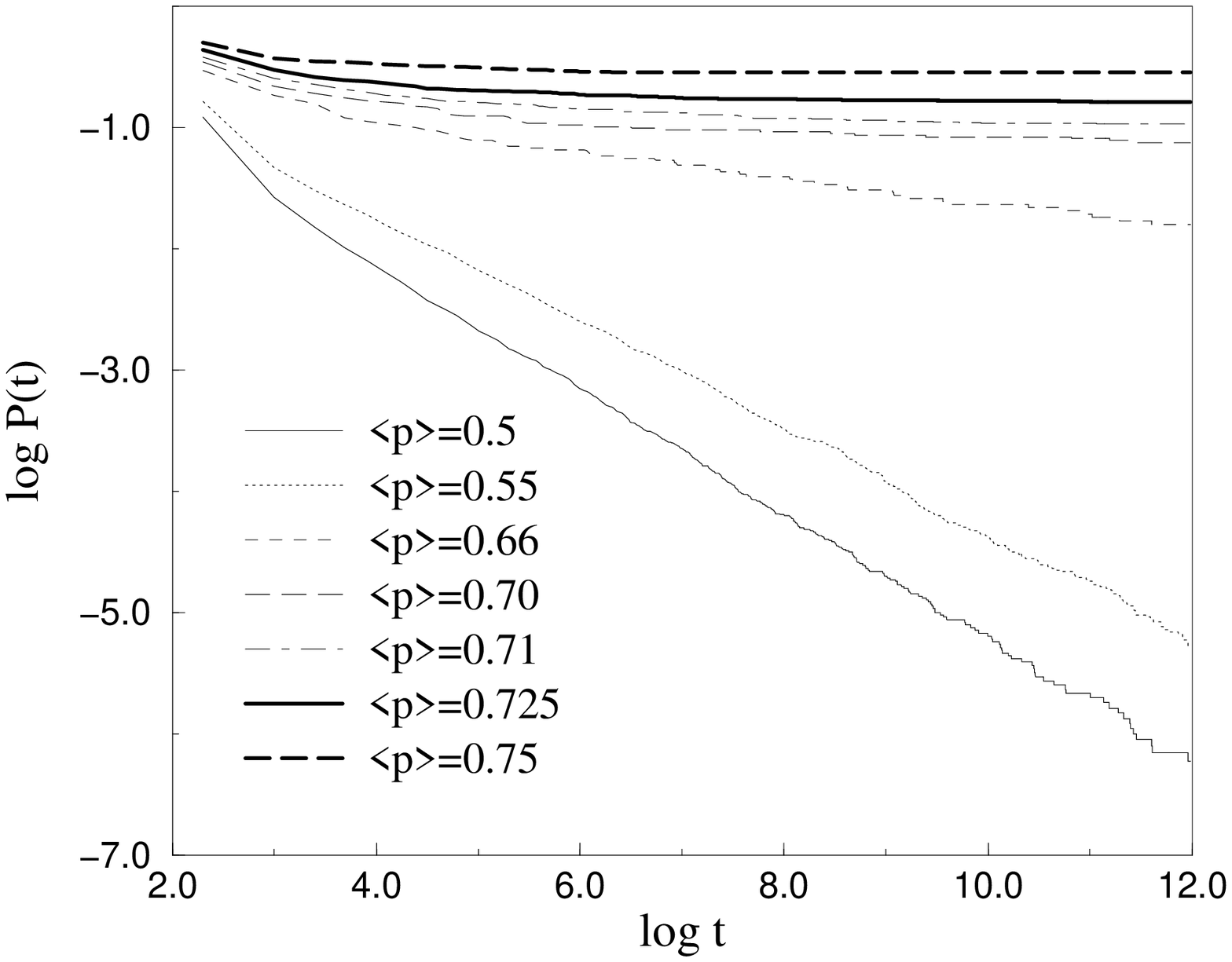,width=6cm}}
\caption{Surviving probability for  spreading experiments, with
different values of $<p>$, 
as a function of time for the disordered model (upper plot), and for the
non-Markovian model (lower plot).
}
\end{figure}

\begin{figure}
\caption{ Averaged square distance from the initial seed, with
different values of $<p>$, in a spreading experiment
as a function of time for the disordered model (upper plot), and for the
non-Markovian model (lower plot).
}
\end{figure}



\begin{references}


\bibitem{conjecture} P. Grassberger, Z. Phys. B {\bf 47}, 365 (1982);
H.K. Janssen, Z. Phys. B {\bf 42}, 151 (1981).

\bibitem{Kin} {\it Percolation Structures and
Processes}, edited by G. Deutscher, R. Zallen, and J. Adler,
Annals of the Israel Physical Society, Vol. 5 (Adam Hilger, Bristol, 1983).

 \bibitem{Marro}
J. Marro and R. Dickman,  {\it Nonequilibrium Phase Transitions and Critical
Phenomena}, Cambridge University Press, (Cambridge, 1996).
               
\bibitem{GG} G. Grinstein, Z.-W. Lai and D.A. Browne, Phys. Rev.
A {\bf 40}, 4820 (1989). 


\bibitem{inf}  M. A. Mu\~noz, G. Grinstein,
R. Dickman and R. Livi,
Phys. Rev. Lett. {\bf 76}, 451, (1996). See also  M. A. Mu\~noz, G. Grinstein,
R. Dickman and R. Livi,
 Physica {\bf D 103}, 485 (1997).


\bibitem{CP} T. M. Ligget,  {\it Interacting Particle Systems},
(Springer Verlag, New York, 1985).
                                                    

\bibitem{catal} R. M. Ziff, E. Gulari, and Y. Barshad, Phys. Rev. Lett.
{\bf 56}, 2553 (1986); I. Jensen, H.C. Fogedby and R. Dickman,
Phys. Rev. {\bf 41}, 3411 (1990).


\bibitem{bawodd} H. Takayasu and A. Yu. Tretyakov, Phys. Rev. Lett.
{\bf 68}, 3060 (1992); I. Jensen, Phys. Rev. E {\bf 47}, 1 (1993);
 I. Jensen, J. Phys. {\bf A 26}, 3921 (1993).                       

\bibitem{RFT} H.D.I. Abarbanel and J.B. Bronzan, Phys. Rev.
D {\bf 9}, 2397 (1974).

\bibitem{Kinzel} W. Kinzel, Z. Phys. {\bf B 58}, 229 (1985).

\bibitem{Noest1} A. J. Noest, Phys. Rev. Lett. {\bf 57}, 91 (1986).

\bibitem{Noest2} A. J. Noest, Phys. Rev. {\bf B 38}, 2715 (1988).

\bibitem{Harris}  A. B. Harris,  J. Phys. {\bf C 7}, 1671 (1974).

\bibitem{Grif} R. B. Griffiths Phys. Rev. Lett. {\bf 23}, 17 (1969).
 
\bibitem{ron1} A. G. Moreira and R. Dickman, Phys. Rev.
{\bf E 54}, R3090 (1996).

\bibitem{ron2} R. Dickman and A. G. Moreira, Cond-mat/9709082.


\bibitem{Iwan} I. Jensen, Phys. Rev. Lett. {\bf 77}, 4988 (1996).

\bibitem{Janssen} H. K. Janssen, Phys. Rev. {\bf E 55}, 6253 (1997).


\bibitem{Obukhov} S. P. Obukhov,
 Pis'ma Zh. Eksp. Teor. Fiz. {\bf 45}, 139 (1987)
[JEPT Lett {\bf 45}, 172 (1987)].


\bibitem{Bramson} M. Bramson, R. Durret and R. H. Schnmann,
Ann. Prob. {\bf 19}, 960 (1991).
 
\bibitem{rts}  L. Pietronero and W. R. Schneider, Physica {\bf A 119},
249 (1989); see also M. Marsili, J. Stat. Phys. {\bf 77}, 733 (1994);
and A. Gabrielli, M. Marsili, R. Cafiero and L. Pietronero,
 J. Stat. Phys. {\bf 84}, 889 (1996).


\bibitem{torre}  P. Grassberger and A. De La Torre,
Ann. Phys. {\bf 122}, 373 (1979).

\bibitem{qdbm} R. Cafiero, A. Gabrielli, M. Marsili, L. Pietronero and
L. Torosantucci, {\em Phys. Rev. Lett.} {\bf 79}, 1503 (1997).

\bibitem{rts1} M. Marsili, {\em Europhys. Lett.} {\bf 28}, 385
(1994); Cafiero, A. Gabrielli, M. Marsili and L. Pietronero, 
{\it Phys. Rev.} {\bf  E 54}, 1406 (1996).


\bibitem{Yuhai}  G. Grinstein, M. A. Mu\~noz and Y. Tu,
Phys. Rev. E. {\bf 56}, 5101 (1997).


\bibitem{Matteo} M. Vendruscolo and M. Marsili, Phys. Rev. {\bf E 54}, R1021
(1996).



\bibitem{Feller} W. Feller {\it An Introduction to Probability Theory and its
Applications}, (Wiley, New York, 1968).


\bibitem{otras} The same technique can be applied to other probability   
distributions without additional problems.
                                       

\bibitem{aging} A recent review can be found in:
E. Vincent, J. Hammann, M. Ocio, J. P. Bouchaud and
L.F. Cugliandolo, in {\it Complex behaviour of glassy
systems},  ed. M. Rub{\'\i}, Lecture Notes in Physics, {\bf 492}, 184 (1996).

\bibitem{note1} In pure systems, an initial seed spreads
asymptotically  in the active zone with a law $R^2 \propto t^2$.
In $d=2$ strong evidence has been found that no sublinear regime
exists \cite{ron2}.


\end{references}
\end{document}